\documentstyle[12pt,epsfig]{article}

\newlength{\dinwidth}
\newlength{\dinmargin}
\begin{document}
\def\bold#1{\setbox0=\hbox{$#1$}%
     \kern-.025em\copy0\kern-\wd0
     \kern.05em\%\baselineskip=18ptemptcopy0\kern-\wd0
     \kern-.025em\raise.0433em\box0 }
\def\slash#1{\setbox0=\hbox{$#1$}#1\hskip-\wd0\dimen0=5pt\advance
         to\wd0{\hss\sl/\/\hss}}
\newcommand{\be}{\begin{equation}}
\newcommand{\ee}{\end{equation}}
\newcommand{\bea}{\begin{eqnarray}}
\newcommand{\eea}{\end{eqnarray}}
\newcommand{\nn}{\nonumber}
\newcommand{\dd}{\displaystyle}
\newcommand{\bra}[1]{\left\langle #1 \right|}
\newcommand{\ket}[1]{\left| #1 \right\rangle}
\newcommand{\spur}[1]{\not\! #1 \,}
\thispagestyle{empty}
\vspace*{1cm}
\rightline{BARI-TH/02-440}
\vspace*{2cm}
\begin{center}
  \begin{LARGE}
  \begin{bf}
$B^- \to K^- \chi_{c0}$ decay\\ from charmed meson rescattering

\vspace*{0.5cm}
  \end{bf}
  \end{LARGE}
\end{center}
\vspace*{8mm}
\begin{center}
\begin{large}
P. Colangelo$^a$, F. De Fazio$^a$ and T.N. Pham$^b$
\end{large}
\end{center}
\begin{center}
\begin{it}
$^a\;\;$Istituto Nazionale di Fisica Nucleare, Sezione di Bari, Italy\\
$^b\;\;$ Centre de Physique Th\'eorique, \\
Centre National de la Recherche Scientifique, UMR 7644\\
\'Ecole Polythechnique, 91128 Palaiseau Cedex, France\\
\end{it}
\end{center}
\begin{quotation}
\vspace*{1.5cm}
\begin{center}
  \begin{bf}
  Abstract\\
  \end{bf}
\end{center}
\vspace*{0.5cm}
\noindent
We study the process $B^- \to K^- \chi_{c0}$ considering  intermediate
charmed meson rescattering effects. For this decay mode the naive
factorization ansatz would predict a vanishing amplitude.
We estimate  contributions from the $D^{(*)}_{s} D^{(*)}\to K^- \chi_{c0}$
rescattering amplitudes, and compare the result with
recent experimental measurements. We find that rescattering effects
are able to produce a large branching ratio consistent with
measurements by Belle Collaboration.
We also consider rescattering effects in $B^- \to K^- J/\psi$,
arguing that they play a similar role in producing a large branching fraction
for this colour-suppressed decay mode.

\vspace*{0.5cm}
\end{quotation}
\newpage
\baselineskip=18pt
\vspace{2cm}
\noindent
\section{Introduction}
Understanding strong interaction effects in weak exclusive
heavy hadron decays is of great importance to gain information
on fundamental aspects of strong and weak interaction phenomenology,
both in the Standard Model and beyond.  In this respect
the factorization ansatz, that allows a treatment of nonleptonic
decay amplitudes
by factorizing hadronic matrix elements of four-quark operators as products
of two current matrix elements, has been a widely used working tool
for the analysis of $B$ decays to charmed and charmless hadrons in the
final state. In those decays in which the effective Wilson coefficients
are not colour suppressed and the tree-level $(V-A)\times (V-A) $
current matrix elements do not vanish, it is found that factorization
provides a reasonable description of data  \cite{Neubert:1997uc}.
However, the recent
observation of the decay mode $B^- \to K^- \chi_{c0}$, reported
by the Belle Collaboration together with the measurement of the branching
fraction \cite{Abe:2002mw}:
\be
{\cal B}(B^- \to K^- \chi_{c0})=(6.0^{+2.1}_{-1.8}\pm 1.1) \times 10^{-4}
\;\;\;, \label{belledatum}
\ee
demonstrates the inadeguacy of the factorization model
in the calculation of nonleptonic $B$ decay amplitudes
for colour suppressed $B$ to charmonium transitions. A large
nonfactorizable term is needed  to account for the observed branching ratio.
As a matter of fact, the result (\ref{belledatum})
implies that the rate of $B$ decays into a kaon and
the $0^{++}$ state of the charmonium system, $\chi_{c0}$,
is comparable with the $B$ decay rate
into a kaon and $J/\psi$, and indeed  the measurement
of the ratio of the two branching fractions,
reported by the same Collaboration, is:
\be
{{\cal B}(B^- \to K^- \chi_{c0}) \over
{\cal B}(B^- \to K^- J/\psi)} =(0.60^{+0.21}_{-0.18}\pm 0.05\pm0.08) \;\;\;.
\label{belleratio}
\ee
The experimental results (\ref{belledatum}) and (\ref{belleratio})  are in
conflict with the vanishing of
the amplitude of $B \to K \chi_{c0}$ computed by the factorization
ansatz, while the amplitude governing $B \to K J/\psi$ is different
from zero in the same approximation. This can be easily shown:
the effective Hamiltonian
governing both the transitions \footnote{We neglect the $B^-$
annihilation transition, which is governed by the CKM matrix element
$V_{ub}$.}:
\bea
H_W={G_F \over \sqrt 2} &\Big\{&V_{cb} V^*_{cs}
\Big( c_1(\mu) {\cal O}_1(\mu)+ c_2(\mu) {\cal O}_2(\mu) \Big)
\nonumber \\
&-&V_{tb} V^*_{ts} \sum_i c_i(\mu) {\cal O}_i(\mu) \Big \} + h.c.
\label{hamiltonian}
\eea
involves only vector and axial-vector $\bar c c$ operators:
\bea
{\cal O}_1&=&(\bar c b)_{V-A} (\bar s c)_{V-A}  \nonumber \\
{\cal O}_2&=&(\bar s b)_{V-A} (\bar c c)_{V-A} \nn \\
{\cal O}_{3(5)}&=&(\bar s b)_{V-A} \sum_q(\bar q q)_{V-A [V+A]} \nn \\
{\cal O}_{4(6)}&=&(\bar s_i b_j)_{V-A} \sum_q(\bar q_j q_i)_{V-A [V+A]} \\
{\cal O}_{7(9)}&=&{3\over 2}(\bar s b)_{V-A} \sum_q e_q (\bar q q)_{V+A [V-A]}
\nn \\
{\cal O}_{8(10)}&=&{3\over 2}(\bar s_i b_j)_{V-A} \sum_q e_q (\bar q_j q_i)_{V+A[V-A]} \nonumber \label{effectiveoper}
\eea
($i,j$ are color indices and
$(\bar q q)_{V\mp A}= \bar q \gamma^\mu (1 \mp \gamma_5) q$),
and therefore the factorized amplitude
\begin{equation}
{\cal A}_F(B^- \to K^- \chi_{c0}) ={G_F \over \sqrt 2} V_{cb} V^*_{cs}
\Big[a_2+\sum_{i=3,5,7,9} a_i \Big]
\langle K^-|(\bar s b)_{V-A}|B^-\rangle
\langle \chi_{c0} |(\bar c c)_{V \mp A}|0\rangle \label{matrixel}
\end{equation}
vanishes since the current matrix elements
$\langle \chi_{c0} |(\bar c c)_{V,A}|0\rangle$ are zero.
Instead, the experimental result  (\ref{belledatum}) corresponds to
${\cal A}_{exp}(B^- \to K^- \chi_{c0})=(3.39\pm0.68)\times 10^{-7}$ GeV.
On the other hand in the case of $B \to K J/\psi$,
since $\langle J/\psi |(\bar c c)_{V}|0\rangle \ne 0$,
a nonvanishing factorized amplitude,  analogous to
(\ref{matrixel}), can be obtained
once the combinations of Wilson coefficients $a_2=c_2+c_1/N_c$ and
$a_i=c_i+c_{i+1}/N_c$,\footnote{The definition of $a_7$ and $a_9$
includes a factor $e$.} and the matrix element
$\langle K^-|(\bar s b)_{V-A}|B^-\rangle$ are provided.

Corrections to naive factorization involve gluon exchanges between
the charmonium system and the quarks in  $B$ and $K$ mesons. For a
class of nonleptonic $B \to M_1 M_2$ decays it has been argued
that, in the large $m_b$ limit, non factorizable corrections are
dominated by hard (perturbatively calculable) gluon exchanges,
while soft effects are confined to the $(B, \, M_1)$ system, where
$M_1$ is the meson picking up the spectator quark in $B$ decay.
This is the case of several processes where $M_1$ and $M_2$ are
light mesons. However, when the meson which does not pick up the
spectator quark is heavy, such a result no longer holds
\cite{bbns}. In order to apply the QCD-improved factorization
model to $B$ decays to charmonium plus a kaon, either the $c {\bar
c}$ state should be considered light with respect to the $B$
meson, or one has to invoke the small transverse size of the $\bar
c c$ system in order to assume a tiny overlap of the quarkonium
wave function with the kaon wave function. However, an  analysis
of non factorizable corrections due to hard gluon exchanges in $B
\to K \chi_{c0}$ has revealed the presence of infrared
singularities, showing a difficulty of the method when applied to
this decay mode \cite{Song:2002mh}.
In the present note we investigate another effect, namely
the $K \chi_{c0}$  production by rescattering of open charm mesons
$D_s^{(*)} D^{(*)}$ etc. primarily produced in $B^-$ decays.
The corresponding amplitude is mainly obtained by the operators
${\cal O}_{1}$ and ${\cal O}_{2}$ in (\ref{hamiltonian}), and therefore
this process could produce a sizeable contribution to
$B \to K \chi_{c0}$ owing to the relatively large values of the
corresponding Wilson coefficients $c_1$ and $c_2$.
Analogous effects were investigated in
\cite{Colangelo:1989gi}, and have recently received new attention
\cite{Ciuchini:2001gv,Isola:2001ar}.

Rescattering of intermediate $D_s^{(*)} D^{(*)}$ mesons can also contribute
to the transition $B \to K J/\psi$ and therefore we also consider this decay
mode. Our conclusion is that, although the calculation presents uncertainties
the size of which we shall try to assess, rescattering effects represent a
non-negligible contribution to the decay channels $B^- \to K^- \chi_{c0}$
and $B^- \to K^- J/\psi$.

\section{Process $B^- \to D_s^{(*)-} D^{(*)0}\to K^- \chi_{c0}$}

In the charm sector, rescattering effects
have been recognized as a source of sizeable contributions
in hadronic $D$ meson decays. As the mass of the
decaying $B$ meson is larger than the $D$ meson mass, one could
suppose a minor role of such processes in $B$ transitions,
since one naively expects 
that high momentum final state particles move fast away from the
interaction region without having the possibility to rescatter
\cite{Bjorken:kk}. However, in a number of analyses
it has been shown that rescattering effects can play an important role even
in $B$ decays
\cite{Gronau:1997an,Neubert:1997wb,Gronau:1998gr,Falk:1998wc}.

\begin{figure}[ht]
\begin{center}
\mbox{\epsfig{file=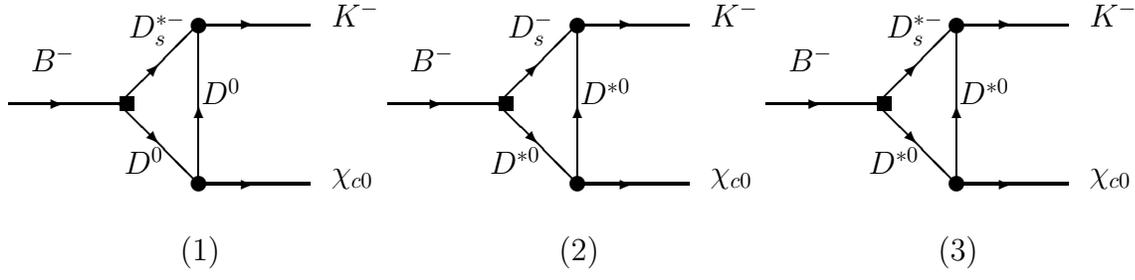, width=15cm}}
\vspace*{-0.5cm}
\end{center}
\caption{\baselineskip 15pt
Diagram contributing to the decay
$B^- \to K^- \chi_{c0}$. The boxes represent weak vertices,
the dots strong couplings}
\vspace*{1.0cm}
\label{diagrams}
\end{figure}

It is worth attempting an estimate of the size of rescattering effects
in color-suppressed $B$ decays to final states containing heavy
particles. We  concentrate on two-body charmed meson
contributions \footnote{The role of  inelastic effects in $B$ decays has
been emphasized in \cite{Donoghue:1996hz}.},
which can be included through a number of dynamical assumptions.
We consider a set of amplitudes corresponding to the diagrams in
fig.\ref{diagrams}, that represent $t$-channel contributions to the
final state interaction. The charmed intermediate states
$D_s^{(*)}$ and $D^{(*)}$ rescatter to $\chi_{c0}$ and $K$ by the
exchange of one resonance states, $D$ and  $D^{*}$. We treat the
exchanged resonances as virtual particles,
with their propagators taken as Breit-Wigner forms.

The analysis of the diagrams in fig.\ref{diagrams} involves the
weak matrix elements governing the transitions
$B^- \to D_s^{(*)-} D^{(*)0}$, and the strong couplings
between the charmed states $D_s^{(*)-} D^{(*)0}$ and the kaon and $\chi_{c0}$.
There is experimental evidence that the calculation of the amplitude
by factorization reproduces the main
features of the  $B^- \to D_s^{(*)-} D^{(*)0}$
decay modes \cite{Luo:2001mc}. Therefore, neglecting the contribution
of the operators ${\cal O}_{3-10}$ in (\ref{effectiveoper}), we can write:
\begin{equation}
\langle D_s^{(*)-} D^{(*)0} | H_W | B^- \rangle =
\displaystyle{G_F \over \sqrt{2}}V_{cb}V_{cs}^* a_1
\langle D^{(*)0} | (V-A)^\mu | B^- \rangle
\langle D_s^{(*)-}| (V-A)_\mu | 0 \rangle
\label{fact}
\end{equation}
with $a_1=c_1+c_2/N_c$.
Using the heavy quark effective theory, the above matrix elements
can be expressed in terms of a single form factor, the Isgur-Wise
function $\xi$, and  a single leptonic constant $\hat F$ \cite{DeFazio:2000up}.
This can be shown expressing the fields $H_a$
describing the negative parity  $J^P=(0^-,1^-)$ ${\bar q }Q$ meson doublet
\begin{equation}
H_a = \frac{(1+{\rlap{v}/})}{2}[P_{a\mu}^*\gamma^\mu-P_a\gamma_5] \;\;,
\label{neg}
\end{equation}
in terms of   operators $P^{*\mu}_a$ and $P_a$ respectively annihilating the
$1^-$  and $0^-$  mesons of four-velocity $v$ ($a=u,d,s$ is a light flavour
index),
and writing the $B^- \to D^{(*)0}$ matrix elements as follows:
\begin{eqnarray}
<D^0(v^\prime)|V^\mu|B^-(v)>&=&\sqrt{m_B m_D} \; \xi(v \cdot v^\prime)
(v+v^\prime)^\mu\nonumber \\
<D^{*0}(v^\prime,\epsilon)|V^\mu|B^-(v)>&=& - i
\sqrt{m_B m_{D^*}} \; \xi(v \cdot v^\prime) \; \epsilon^*_{\beta} \;
\varepsilon^{\alpha \beta \gamma \mu} v_\alpha v^\prime_\gamma \label{BD} \\
<D^{*0}(v^\prime,\epsilon)|A^\mu|B^-(v)>&=&
\sqrt{m_B m_{D^*}} \; \xi(v \cdot v^\prime) \; \epsilon^*_{\beta}
[(1+v \cdot v^\prime) g^{\beta \mu}- v^\beta v^{\prime\mu}] \,\,\, .\nonumber
\end{eqnarray}
In (\ref{BD}) $\epsilon$ is the $D^*$ polarization vector and
$\xi(v \cdot v^\prime)$ represents the Isgur-Wise form factor.

The weak current for the transition from a heavy to a light quark
$Q \to q_a$, given at the quark level by
$\bar q_a \gamma^\mu(1-\gamma_5) Q$, can be written in terms of a heavy meson
and light pseudoscalars.
The octet of the light pseudoscalar mesons is represented by
 $\displaystyle \xi=e^{i {\cal M} \over f}$, with
\begin{equation}
{\cal M}=
\left (\begin{array}{ccc}
\sqrt{\frac{1}{2}}\pi^0+\sqrt{\frac{1}{6}}\eta & \pi^+ & K^+\nonumber\\
\pi^- & -\sqrt{\frac{1}{2}}\pi^0+\sqrt{\frac{1}{6}}\eta & K^0\\
K^- & {\bar K}^0 &-\sqrt{\frac{2}{3}}\eta
\end{array}\right )
\end{equation}
and $f\simeq f_{\pi}=131 \; MeV$, and the effective heavy-to-light current,
written at the lowest order in the light meson derivatives, reads:
\begin{equation}
L^\mu_a =
{\hat F \over 2} Tr[\gamma^\mu (1- \gamma_5) H_b \xi^\dagger_{ba}]
\,\,\, .
\end{equation}
In this way, all the matrix elements
$<0|\bar q_a \gamma^\mu (1-\gamma_5) c|D_a^{(*)}(v)>$
are related to the single constant $\hat F$:
\begin{eqnarray}
<0|\bar q_a \gamma^\mu \gamma_5 c|D_a(v)> &=&f_{D_a} m_{D_a} v^\mu \nonumber
\\
<0|\bar q_a \gamma^\mu c|D^*_a(v,\epsilon)> &=&f_{D^*_a} m_{D^*_a}
\epsilon^\mu
\end{eqnarray}
with $f_{D_a}=f_{D^*_a}={\hat F \over \sqrt{m_{D_a}}}$.

Other hadronic quantities appearing in the diagrams in fig.\ref{diagrams}
are  the strong couplings $D_s^{(*)} D^{(*)} K$ and
$D^{(*)} D^{(*)} \chi_{c0}$.
The $D_s^{(*)} D^{(*)} K$ couplings, in the soft $\vec p_K \to 0$ limit,
can be related to a single effective constant $g$, as it turns out
considering the effective QCD Lagrangian describing the
strong interactions between the heavy $D^{(*)}_a D^{(*)}_b$ mesons and
the octet of the light pseudoscalar mesons \cite{hqet_chir}:
\begin{equation}
{\cal L}_I = i \;
g \; Tr[H_b \gamma_\mu \gamma_5 {A}^\mu_{ba} {\bar H}_a]  \label{L}
\end{equation}
with the operator $A$ in (\ref{L}) given by
\begin{equation}
{A}_{\mu ba}=\frac{1}{2}\left(\xi^\dagger\partial_\mu \xi-\xi
\partial_\mu \xi^\dagger\right)_{ba} \; .
\label{chirallag}
\end{equation}
This allows to relate the $D_s^{(*)} D^{(*)} K$ couplings,
defined through the matrix elements
\begin{eqnarray}
<D^0(p) K^-(q)|D_s^{*-} (p+q,\epsilon))>&=&
g_{D_s^{*-} D^0 K^-} \, \, (\epsilon \cdot q) \nonumber\\
<D^{*0}(p,\eta) K^-(q)|D_s^{*-} (p+q,\epsilon))>&=& i \,\,
\epsilon^{\alpha \beta \mu \gamma} \, p_\alpha \, \epsilon_\beta
\, q_\mu \eta^*_\gamma \,\, g_{D_s^{*-} D^{*0} K^-}  \label{gddk}
\end{eqnarray}
to the effective coupling $g$:
\begin{eqnarray}
g_{D_s^{*-} D^{0} K^-}&=& - 2 \sqrt{m_D m_{D_s^*}}
{\displaystyle g \over f_K} \nn \\
g_{D_s^{*-} D^{^*0} K^-}&=& 2 \sqrt{m_{D_s^*} m_{D^*}}
{\displaystyle g  \over f_K} \,\,\,\ .
\label{gcoupl}
\end{eqnarray}

As for the coupling of the $\chi_{c0}$ state
to a pair of $D$ mesons, defined by the matrix element:
\be
\langle D^{0}(p_1) \bar D^{0}(p_2)| \chi_{c0} (p)\rangle= g_{D D \chi_{c0}}
\,\,\, ,\label{gddchi}
\ee
an estimate can be obtained
considering the $D$ matrix element of the scalar $\bar c c$ current:
$\displaystyle \langle D(v^\prime) | \bar c c | D(v) \rangle$,
assuming the dominance
of the nearest resonance, i.e. the scalar $\bar c c$ state,
in the $(v-v^\prime)^2$-channel and using the normalization
of the Isgur-Wise form factor at the zero-recoil point
$v=v^\prime$. This allows us to express  $g_{D D \chi_{c0}}$
in terms of the constant  $f_{\chi_{c0}}$ that parameterizes the matrix
element
\begin{equation}
\langle 0| \bar c c | \chi_{c0}(q)\rangle= f_{\chi_{c0}} m_{\chi_{c0}} \,\,\, .
\label{fchi}
\end{equation}
The method can also be applied to $g_{D^* D^* \chi_{c0}}$. One obtains:
\begin{equation}
g_{D D \chi_{c0}}= - 2  {\displaystyle m_D m_{\chi_{c0}} \over f_{\chi_{c0}}}
\hspace*{1cm}
g_{D^* D^* \chi_{c0}} = 2  {\displaystyle m_{D^*} m_{\chi_{c0}}
\over f_{\chi_{c0}}} \,\,\,.
\label{gchi}
\end{equation}

It is worth noticing, however, that the determinations of the couplings
described  above do not account for
the off-shell effect of the exchanged  $D$ and $D^*$
particles,  the virtuality of which can be large.
As discussed in the literature, a method to account for such effect
relies on the introduction of form factors:
\begin{equation}
g_i(t)=g_{i0}\,F_i(t)\,,
\label{offshell}
\end{equation}
with $g_{i0}$ the corresponding on-shell couplings
(\ref{gddk}), (\ref{gddchi}). A simple pole representation for
$F_i(t)$ is:
$F_i(t)=\displaystyle{\Lambda_i^2 -m^2_{D^{(*)}} \over \Lambda_i^2 -t}$,
consistent with QCD counting rules
\cite{Gortchakov:1995im}. The parameters in the form factors represent
a source of uncertainty in our analysis.

We have the elements for computing the diagrams in fig.\ref{diagrams}.
The absorptive part of the amplitude (1) simply reads:
\begin{equation}
{\rm Im} \, A_1 = {\sqrt{\lambda(m_B^2,m^2_{D^*_s},m^2_{D})} \over 32 \pi m^2_B}
\int_{-1}^{+1} dz {\cal A}(B^- \to D^{*-}_s  D^{0})
{\cal A}(D^{*-}_s  D^{0} \to K^- \chi_{c0})
\end{equation}
with $\lambda$ the triangular function.
Analogous expressions correspond to the diagrams (2) and (3).
Explicitely, the imaginary parts are  given by
\begin{eqnarray}
{\rm Im} \, A_1 &=& {K f_{D^*_s} m_{D^*_s} \sqrt{m_B m_D} (m_B+m_D)
\over 32 \pi m_B^2 m_D}
\lambda^{1/2}(m_B^2,m_{D^*_s}^2,m_D^2)\int_{-1}^1 dz \, g_{D^*_s D
K}(t) g_{DD\chi}(t) \nonumber \\
&&\xi \left( {m_B^2-m_{D^*_s}^2+m_D^2 \over 2 m_B m_D} \right)
{-q^0+\displaystyle{k^0 q \cdot k \over m_{D^*_s}^2} \over
t-m_D^2} \nonumber \\ \nonumber \\
{\rm Im} \, A_2 &=& {K f_{D_s}  \sqrt{m_B m_{D^*}} \over 32 \pi m_B^2 }
\lambda^{1/2}(m_B^2,m_{D_s}^2,m_{D^*}^2)\int_{-1}^1 dz \, g_{D_s D^*
K}(t) g_{D^*D^*\chi}(t) \nonumber \\
&&\xi \left( {m_B^2-m_{D_s}^2+m_{D^*}^2 \over 2 m_B m_D^*} \right)
\Big\{ \left[ {m_K^2-q \cdot k \over m_{D_s}^2 } -1 \right] [-(1+v
\cdot v_D) q \cdot k + v_D \cdot k (q^0+v_D \cdot q)]
\nonumber \\ \nonumber \\
&-&{m_K^2-q \cdot k \over m_{D_s}^2 }[-(1+v \cdot v_D) m_{D_s}^2 +
v_D \cdot k (k^0+v_D \cdot k)] \Big\} \nonumber \\ \nonumber \\
{\rm Im} \, A_3 &=& - {K f_{D^*_s} m_{D^*_s} \sqrt{m_B m_D^*} \over 16
\pi m_B^2 } \lambda^{1/2}(m_B^2,m_{D^*_s}^2,m_{D^*}^2)\int_{-1}^1
dz \, g_{D^*_s D^*
K}(t) g_{D^*D^*\chi}(t) \nonumber \\
&&\xi \left( {m_B^2-m_{D^*_s}^2+m_{D^*}^2 \over 2 m_B m_D^*}
\right) {q^0 v_D \cdot k -k^0 v_D \cdot q  \over t-m_{D^*}^2}\,,
\label{impartschi}
\end{eqnarray}
where: $K=\displaystyle{G_F \over \sqrt{2}}V_{cb} V_{cs}^* a_1$,
$q^0=\displaystyle{m_B^2+m_K^2-m_\chi^2 \over 2 m_B}$, $|\vec
q|=\displaystyle{\lambda^{1/2}(m_B^2,m_K^2, m_\chi^2) \over 2
m_B}$, together with:
$\displaystyle v \cdot v_D={m_B-k^0 \over m_{D_i}}$,
$\displaystyle v_D \cdot k = {m_B k^0-m_{D_{s,i}}^2 \over m_{D_i}}$,
$\displaystyle v_D \cdot q = {m_B q^0- q \cdot k \over m_{D_i}}$,
$\displaystyle k^0= {m_B^2+m_{D_{s,i}}^2-m_{D_i}^2 \over 2 m_B }$,
$\displaystyle |\vec k|= {\lambda^{1/2}(m_B^2,m_{D_{s,i}}^2,m_{D_i}^2) \over
2m_B}$
and $\displaystyle m_{D_{s,1}}=m_{D_{s,3}}=m_{D^*_s}$, $m_{D_{s,2}}=m_{D_s}$;
$\displaystyle m_{D_1}=m_D$, $\displaystyle m_{D_2}=m_{D_3}=m_{D^*}$.

The dispersive parts of the  amplitudes
in fig.\ref{diagrams} can be estimated using
\begin{equation}
{\rm Re} {\cal A}_i (m_B^2) = {1 \over \pi} PV \int_{s_{th}^{(i)}}^{+\infty}
d s^\prime { Im {\cal A}_i (s^\prime) \over s^\prime - m_B^2}
\end{equation}
with the thresholds $s_{th}^{(i)}$ given by: $s_{th}^{(1)}=(m_{D^*_s}+m_D)^2$,
$s_{th}^{(2)}=(m_{D_s}+m_{D^*})^2$ and $s_{th}^{(3)}=(m_{D^*_s}+m_{D^*})^2$,
respectively.
It can be assumed that such expressions are dominated by the region
close to the pole $m_B^2$. Therefore, we compute the integrals,
that in general depend on the asympotic behavior of the spectral
functions $Im {\cal A}_i (s^\prime)$, by using
a cutoff not far from the $B$ meson mass, chosen in the range $35-40$ GeV$^2$.

A comment on other contributions to $K^- \chi_{c0}$
via final state rescattering is in order, since also the
$D^{(*)}$ and  $D^{(*)}_s$ 
excitations could be considered as intermediate
states. These terms are  suppressed by smaller values of the
universal form factors and of the leptonic decay constants; therefore,
the amplitudes in fig.\ref{diagrams} represent
the main contributions that need to be analyzed.

\section{Mode $B^- \to K^- J/\psi$}

Before attempting a numerical estimate of the amplitudes in
fig.\ref{diagrams}, let us consider the decay mode $B^- \to  K^- J/\psi$.
In this case the amplitude obtained in the naive factorization approach,
keeping only the contribution of the operators
${\cal O}_{1}$ and ${\cal O}_{2}$ in (\ref{hamiltonian}), is given by
\begin{eqnarray}
{\cal A}_F(B^- \to K^- J/\psi) &=&2 {G_F \over \sqrt 2} V_{cb} V^*_{cs}
a_2 f_{J/\psi} m_{J/\psi} F_1^{BK}(m^2_{J/\psi}) (\epsilon^* \cdot q)
\nonumber \\
&=& \tilde {\cal A}_F  (\epsilon^* \cdot q) \,\,\,\, ,\label{afactkpsi}
\end{eqnarray}
with the constant  $f_{J/\psi}$ defined by
\be
\langle 0| \bar c \gamma^\mu c | J/\psi(p^\prime,\epsilon)\rangle =
f_{J/\psi} m_{J/\psi} \epsilon^\mu \,\,\,\, ,
\ee
$\epsilon$ the $J/\psi$ polarization vector,
$q$ the kaon momentum and  $F_1^{BK}$ one of the two form factors
parameterizing the matrix element
$\langle K^-| \bar s \gamma^\mu b | B^-\rangle$.

Identifying eq.(\ref{afactkpsi}) with the experimental amplitude
obtained from the measurement
${\cal B}(B^- \to K^- J/\psi)=(1.00 \pm 0.10) \times 10^{-3}$
\cite{pdg}:
$\tilde {\cal A}_{exp}=(1.41 \pm 0.07) \times 10^{-7},$
and using the value $f_{J/\psi}=405\pm14$ MeV, one obtains a result for
the product $|a_2 F_1^{BK}(m^2_{J/\psi})|$. This determination of $a_2$ is
mainly affected by
the uncertainty on $F_1^{BK}$; scanning several form factor models, as done in
\cite{Cheng:1998kd}, one gets $|a_2|=0.2-0.4$, while considering  the
calculation in \cite{Colangelo:1995jv} one obtains $|a_2|=0.38 \pm 0.05$.
Such results are obtained using
$V_{cb}=0.040$ and  $V_{cs}=0.9735$ that correspond to the central values
reported by the Particle Data Group \cite{pdg}.

As discussed at length in the literature, the above values of $a_2$ are
different from the combination $a_2=c_2+c_1/N_c$
of the Wilson coefficients in (\ref{hamiltonian}). As a matter of fact,
from the values $c_1=1.085 (1.109)$ and $c_2=-0.198 (-0.243)$
computed for  ${\overline m}_b(m_b)=4.4$ GeV and
$\Lambda_{\overline{MS}}^{(5)}=290$ MeV in the
naive dimensional regularization (or 't Hooft-Veltman) scheme
\cite{Buras:1998ra},  one would get: $a_2=0.163 (0.126)$.
\footnote{Similar values are obtained varying
${\overline m}_b(m_b)$ and $\Lambda_{\overline{MS}}^{(5)}$.}

\begin{figure}[ht]
\begin{center}
\mbox{\epsfig{file=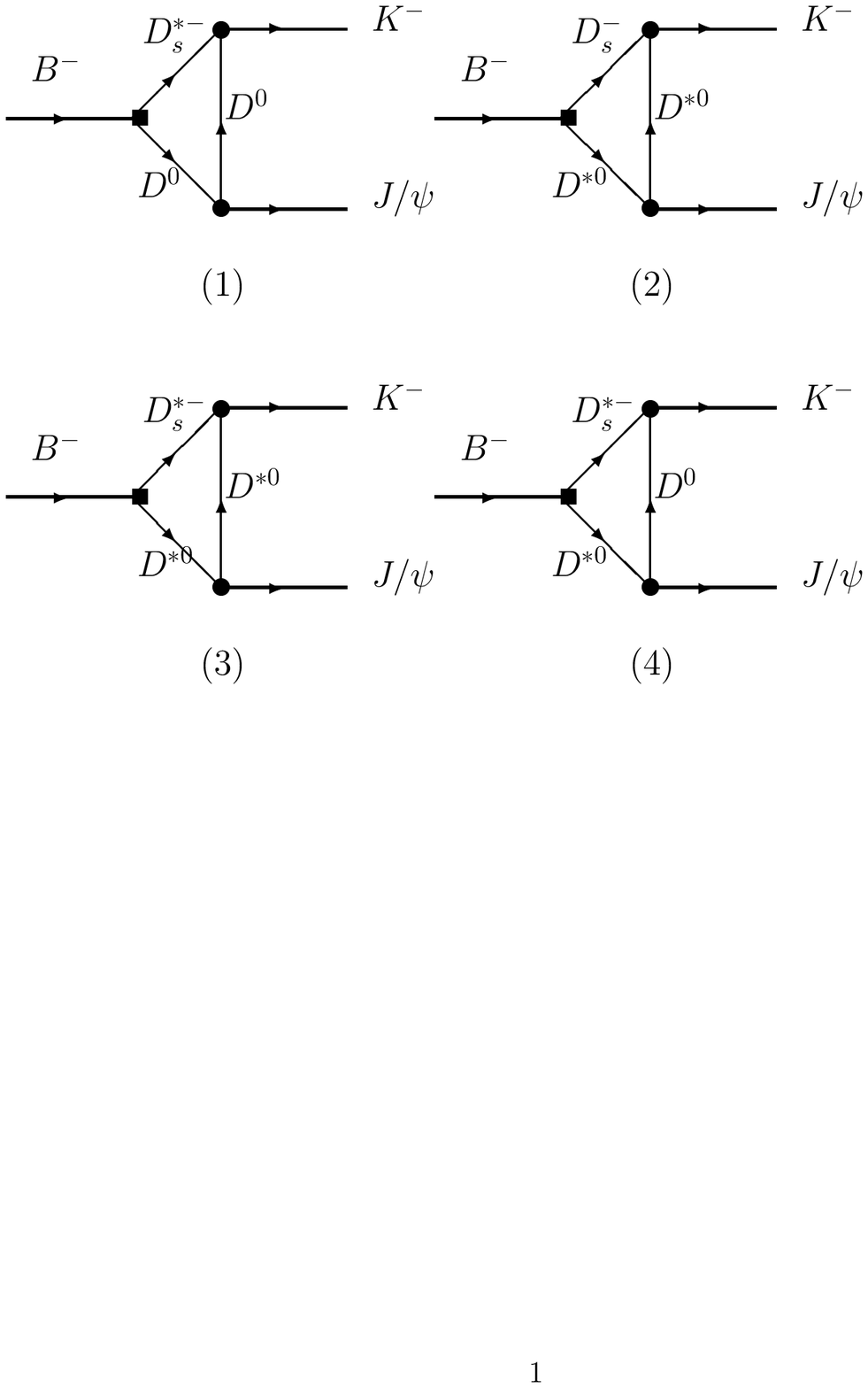, width=8cm}}
\vspace*{-0.5cm}
\end{center}
\caption{\baselineskip 15pt
Rescattering diagrams contributing to $B^- \to K^- J/\psi$.}
\vspace*{1.0cm}
\label{diagrams1}
\end{figure}

Therefore,  nonfactorizable effects are sizeable in $B^- \to K^-
J/\psi$, and indeed in a generalized factorization ansatz $a_2$ is
treated as an effective parameter used to fit the data. The
calculation in the framework of QCD factorization does not allow
to reproduce the fitted value, although an improvement towards the
experimental datum is obtained \cite{Cheng:2000kt}. It is worth
considering rescattering contributions of intermediate charm
mesons, described by diagrams as depicted in fig.\ref{diagrams1}.
The hadronic information for determining such amplitudes are the
same as in Section 2, with the only difference in the  strong
$D^{(*)} D^{(*)} J/\psi$ couplings that can be expressed in terms
of the parameter $f_{J/\psi}$, using the same vector meson
dominance method applied to derive eq. (\ref{gchi}).
%
%
\section{Numerical calculation and discussion}
In order to evaluate the amplitudes in figs.\ref{diagrams},\ref{diagrams1}
we have to fix the values of the various hadronic parameters.
The Wilson coefficient  $a_1$, common to all the amplitudes,
can be put to $a_1=1.0$ as obtained by the
analysis of exclusive $B \to D^{(*)}_s D^{(*)}$ transitions. Moreover,
we use $f_{D_s}=240$ MeV, in the range quoted by the
Particle Data Group \cite{pdg}, and $f_{D^*_s}=f_{D_s}$ consistently
with our approach  that exploits the large $m_Q$ limit.
For the Isgur-Wise universal form factor $\xi$, the expression
$\displaystyle \xi(y)=\Big({ 2 \over y+1}\Big)^2$ is compatible with
the current results from the semileptonic $B \to D^{(*)}$ decays.

A discussion is needed about the $D_s^{(*)} D^{(*)} K$ vertices.
For the effective coupling $g$ in (\ref{gcoupl}) one can use the
CLEO result $g=0.59 \pm 0.01 \pm 0.07$ obtained by the measurement
of  the $D^*$ width \cite{Anastassov:2001cw}. This value is in the
upper side of several theoretical calculations
\cite{Colangelo:2000dp, Colangelo:2002dg}. We choose to be
conservative, and vary this parameter in the range: $0.35 < g <
0.65$ that encompasses the largest part of the predictions.

We use the expression (\ref{gchi})
for the  $D^{(*)} D^{(*)} \chi_{c0}$ vertices,
with $f_{\chi_{c0}}=510\pm40$ MeV obtained by a standard two-point
QCD sum rule analysis.
As for the couplings $D^{(*)} D^{(*)} J/\psi$,
expressions analogous to (\ref{gchi}) involve $f_{J/\psi}$,
for which we use the experimental measurement.
To account for the
off-shell effects of the $D^{(*)}$ exchanged particles,
we use eq.(\ref{offshell}) with two choices for the parameters:
$\Lambda_i=2.5\;\; GeV$ and $\Lambda_i=2.8\;\; GeV$,
corresponding to typical values of the mass
of the radial excitations of $D^{(*)}$ mesons.

The results are reported in Table \ref{table}. One has to notice that the
rescattering amplitudes contribute with different signs to the final result,
with significant cancellations among the various terms.

\begin{table}[h]
\caption{Numerical results for the rescattering amplitudes}
\label{table}
\begin{center}
\begin{tabular}{|| l || c c  || c |} \hline \hline
$B^- \to K^- \chi_{c0}$&${\rm Re} {\cal A}$ (GeV)&${\rm Im} {\cal A}$ (GeV)
& $\Lambda_i$ (GeV) \\\hline
&$-(0.9-1.7)\times 10^{-7}$&$-(0.5-1.0)\times 10^{-7}$&$2.5$\\
&$-(1.4-2.7)\times 10^{-7}$&$-(0.6-1.2)\times 10^{-7}$& $2.8$\\
\hline \hline
$B^- \to K^- J/\psi$&${\rm Re} \tilde {\cal A}$&${\rm Im} \tilde {\cal A}$&$\Lambda_i$ (GeV)\\ \hline
&$(0.1-0.2)\times 10^{-7}$&$-(0.5-0.9)\times 10^{-7}$& $2.5$\\
&$(0.2-0.3)\times 10^{-7}$&$-(0.9-1.7)\times 10^{-7}$&$2.8$\\
\hline \hline

\end{tabular}
\end{center}
\end{table}
%
A few observations are in order. First, in the chosen range of values for
the various
parameters, the rescattering amplitudes are sizeable, and become comparable
to the experimental ones. This observation can be made more quantitative.
Assuming that the amplitude relative to $B^- \to K^- J/\psi$ deviates from
the factorized result because of the contribution of the rescattering term:
$\tilde {\cal A}_{exp} = \tilde {\cal A}_{fact} + \tilde {\cal A}_{resc}$,
one can constrain the values of $\Lambda_i$
for the calculation of  ${\cal B}(B^- \to K^- \chi_{c0})$.
One obtains: $\Lambda_i\simeq 2.7\,\,\, GeV$ and
${\cal B}(B^- \to K^- \chi_{c0})=(1.1-3.5) \times 10^{-4}$, to be compared to
(\ref{belledatum}).  The result seems noticeable, considering the rather
schematic description of the rescattering process.

The second observation is that a correspondence similar to that between
$B^- \to K^- \chi_{c0}$ and $B^- \to K^- J/\psi$ is expected with
analogous decay modes, namely $B^- \to K^- \chi_{c1}$ and
$B^- \to K^- \chi_{c2}$, the mesons $\chi_{c1,2}$ being the axial vector
and the tensor states of the charmonium system. It is worth observing that
$B^- \to K^- \chi_{c2}$ is another process the amplitude of which vanishes in
the naive factorization model; in our approach, we expect a
branching fraction analogous to that of $B^- \to K^- \chi_{c0}$.

We are aware of
the various sources of theoretical uncertainty. The uncertainty related to
the
$B^- \to D_s^{(*)}  D^{(*)}$ vertices can be minimized gaining
experimental information on these decay modes. The
uncertainties in the strong vertices can be reduced by dedicated analyses
using nonperturbative QCD methods (QCD sum rules or lattice QCD).
However, all such uncertainties
only affect the precise numerical predictions and not
our main conclusion. We have found that rescattering amplitudes,
describing a rearrangement
of the quarks in the  final state after the production of pairs of
charmed  mesons, not only cannot be neglected both in
$B^- \to K^- \chi_{c0}$, both in  $B^- \to K^- J/\psi$, but can provide a
large part of the decay amplitudes.
Analogous effects  are expected to be important in similar colour
suppressed
decay modes,  namely  $B^- \to K^- \chi_{c1}$ and $B^- \to K^- \chi_{c2}$.

\end{document}